\documentclass[a4paper,12pt]{article}
\usepackage[latin1]{inputenc}
\usepackage{amssymb}
\title{{\bf  On Decidability Properties of One-Dimensional Cellular Automata  } }
\author{Olivier Finkel  \\{\it   Equipe de Logique Mathématique}
  \\ CNRS et  Universit\'e Paris Diderot Paris 7
 \\ UFR de Mathématiques case 7012, site Chevaleret,\\75205 Paris Cedex 13, 
 France.\\ finkel@logique.jussieu.fr 
}

\date{}
\begin{document}

\newtheorem{The}{Theorem}[section]
\newtheorem{Pro}[The]{Proposition}
\newtheorem{Deff}[The]{Definition}
\newtheorem{Lem}[The]{Lemma}
\newtheorem{Rem}[The]{Remark}
\newtheorem{Exa}[The]{Example}
\newtheorem{Cor}[The]{Corollary}

\newcommand{\vp}{\varphi}
\newcommand{\lb}{\linebreak}
\newcommand{\fa}{\forall}
\newcommand{\Ga}{\Gamma}
\newcommand{\Gas}{\Gamma^\star}
\newcommand{\Gao}{\Gamma^\omega}
\newcommand{\Si}{\Sigma}
\newcommand{\Sis}{\Sigma^\star}
\newcommand{\Sio}{\Sigma^\omega}
\newcommand{\ra}{\rightarrow}
\newcommand{\hs}{\hspace{12mm}

\noi}
\newcommand{\lra}{\leftrightarrow}
\newcommand{\la}{language}
\newcommand{\ite}{\item}
\newcommand{\Lp}{L(\varphi)}
\newcommand{\abs}{\{a, b\}^\star}
\newcommand{\abcs}{\{a, b, c \}^\star}
\newcommand{\ol}{ $\omega$-language}
\newcommand{\orl}{ $\omega$-regular language}
\newcommand{\om}{\omega}
\newcommand{\nl}{\newline}
\newcommand{\noi}{\noindent}
\newcommand{\tla}{\twoheadleftarrow}
\newcommand{\de}{deterministic }
\newcommand{\proo}{\noi {\bf Proof.} }
\newcommand {\ep}{\hfill $\square$}

\maketitle

\begin{abstract}
\noi   In a recent paper Sutner proved that the first-order  theory of the  phase-space  
$\mathcal{S}_\mathcal{A}=(Q^\mathbb{Z}, \longrightarrow)$ of a one-dimensional 
cellular automaton $\mathcal{A}$ whose configurations are elements of 
$Q^\mathbb{Z}$, for a finite set of states $Q$, and where $\longrightarrow$ is the ``next configuration relation", is decidable \cite{Sutner-JCA}. 
He asked whether this result could be extended to a  more expressive logic.  We prove in this paper that this is actuallly the case. We first show that, 
for each one-dimensional cellular automaton $\mathcal{A}$, the phase-space  
$\mathcal{S}_\mathcal{A}$ is an $\om$-automatic structure.  Then,  applying recent results  of Kuske and Lohrey  
 on $\om$-automatic structures,  it follows that  the first-order theory,  extended with some counting and cardinality quantifiers, of the structure 
$\mathcal{S}_\mathcal{A}$, is decidable.  We give some examples of new decidable properties for one-dimensional 
cellular automata. In the case of surjective cellular automata, some more efficient algorithms can be deduced from results of \cite{KuskeLohrey} on 
structures of bounded degree.  On the other hand we show that the case of cellular automata give new results on automatic graphs. 

\end{abstract}

\noi {\small {\bf Keywords:}  One-dimensional cellular automaton; space of configurations; $\om$-automatic structures; first order theory; 
cardinality quantifiers; decidability properties; surjective cellular automaton; automatic graph; reachability relation.}

\section{Introduction}

\noi Some properties of one-dimensional cellular automata, like injectivity or surjectivity, were shown to be decidable by Amoroso and Patt in 
\cite{AmorosoPatt}. These two properties are easily expressed as first-order properties of the  phase-space  
$\mathcal{S}_\mathcal{A}=(Q^\mathbb{Z}, \longrightarrow)$ of a cellular automaton $\mathcal{A}$ whose configurations are elements of 
$Q^\mathbb{Z}$, for a finite set of states $Q$, and where $\longrightarrow$ is the ``next configuration relation". 
\nl It is then very natural to ask whether every first-order property of the structure $\mathcal{S}_\mathcal{A}$ is decidable. Sutner proved recently 
in \cite{Sutner-JCA} that  
this is actually the case, i.e. that ``model-checking for one-dimensional cellular automata in the first order logic $\mathcal{L}(\longrightarrow)$ with 
equality is decidable". He used decidability properties of $\zeta$-B\"uchi  automata reading {\it biinfinite words}. 
\nl On the other hand,  $\om$-automatic structures  are relational  structures 
whose domain and relations are recognizable by B\"uchi  automata reading {\it infinite words}. 
The $\om$-automatic structures
 have very nice decidability and definability properties and have been much studied in the last few years,  
 \cite{Hodgson,BlumensathGraedel04, KNRS}. In particular  
the first-order theory of an $\om$-automatic structure is decidable and the class 
of $\om$-automatic structures is closed under first-order interpretations. 
\nl We show here that, for each one-dimensional cellular automaton $\mathcal{A}$, 
the phase-space $\mathcal{S}_\mathcal{A}=(Q^\mathbb{Z}, \longrightarrow)$ 
is an $\om$-automatic structure. Thus Sutner's result can be deduced from an earlier result of Hodgson in \cite{Hodgson}. 

\hs  But we can now apply to the study of one-dimensional cellular automata some other very recent results on   $\om$-automatic structures. 
Blumensath and Gra\"edel  proved in \cite{BlumensathGraedel04} that  the first-order theory of an $\om$-automatic structure,  
extended with the ``infinity quantifier" $\exists^\infty$, is 
decidable. 
Kuske and Lohrey    proved      in \cite{KuskeLohrey} that the first-order theory,  extended with some counting and cardinality quantifiers, 
of an (injectively)  $\om$-automatic structure   is decidable. These quantifiers are in the form $\exists^\kappa$, for any cardinal $\kappa$,  
or are modulo quantifiers in the form $\exists^{t, k}$, for integers $0\leq t<k$, whose meaning is ``there exist $t$ modulo $k$". 
\nl This gives an answer to Sutner's question who  asked whether his decidability  result could be extended to a  more expressive logic.  

\hs
Sutner's result implies that one can decide some properties like the following ones : ``there exists a $6$-cycle" or ``there exists exactly five fixed points". 
With our new results deduced from properties of  $\om$-automatic structures, 
 we can now decide properties like : ``there exist uncountably many $3$-cycles" or ``there exist infinitely  many $25$-cycles" or ``there exist $t$ modulo $k$ 
$6$-cycles" or ``the set of fixed points is countably infinite", and so on . 

\hs Notice that the algorithms obtained by Sutner in \cite{Sutner-JCA} are in general non elementary. This is also the case for the algorithms obtained by Kuske 
and Lohrey for $\om$-automatic structures. However some more efficient algorithms are obtained in  \cite{KuskeLohrey} in the case of 
$\om$-automatic structures of bounded degree. In the case of one-dimensional cellular automata  this corresponds to the important  class of  surjective 
cellular automata. We give  the complexities of algorithms  obtained from \cite{KuskeLohrey} in this particular case. 

\hs We consider next the set of configurations with finite support of a given cellular automaton, equipped with the next configuration relation.
It is an automatic directed graph, i.e. it is a countable structure presentable with automata reading ``finite words". 
Then we can infer new results on automatic graphs from some results of Sutner \cite{Sutner2}. In particular for 
 each recursively enumerable Turing degree ${\bf d}$ there is an automatic graph in which the 
reachability relation  has exactly the Turing degree ${\bf d}$. This shows the richness of the class of automatic graphs, from the point of view of 
Turing degrees. 

\hs The paper is organized as follows. In Section 2 we recall definitions of cellular automata and of B\"uchi  automata reading infinite words. 
We recall in Section 3 the notion of (injectively)  $\om$-automatic structure. We get some new decidability properties of 
one-dimensional cellular automata in Section 4.  We show in section 5 that the case of cellular automata give new results on automatic graphs.

\section{Cellular automata and  B\"uchi automata}

\noi When $\Si$ is a finite alphabet, a {\it non-empty finite word} over $\Si$ is any 
sequence $x=a_0.a_1\ldots a_k$, where $a_i\in\Sigma$ 
for $i=0, 1,\ldots ,k$ , and  $k$ is an integer $\geq 0$. The {\it length}
 of $x$ is $k+1$, denoted by $|x|$.
 The {\it empty word} has no letter and is denoted by $\lambda$; its length is $0$. 
 For $x=a_0\ldots a_k$, we write $x(i)=a_i$  
and $x[i]=x(0)\ldots x(i)$ for $i\leq k$.
 $\Sis$  is the {\it set of finite words} (including the empty word) over $\Sigma$.
 \nl  The {\it first infinite ordinal} is $\om$.
 An $\om$-{\it word} over $\Si$ is an $\om$ -sequence $a_0.a_1 \ldots a_n \ldots$, where for all 
integers $ i\geq 0$, ~
$a_i \in\Sigma$.  When $\sigma$ is an $\om$-word over $\Si$, we write
 $\sigma =\sigma(0)\sigma(1)\ldots \sigma(n) \ldots $,  where for all $i$,~ $\sigma(i)\in \Si$,
and $\sigma[n]=\sigma(0)\sigma(1)\ldots \sigma(n)$  for all $n\geq 0$.
\nl  
 The {\it set of } $\om$-{\it words} over  the alphabet $\Si$ is denoted by $\Si^\om$.
An  $\om$-{\it language} over an alphabet $\Sigma$ is a subset of  $\Si^\om$.  The complement (in $\Sio$) of an 
$\om$-language $V \subseteq \Sio$ is $\Sio - V$, denoted $V^-$.
\nl A $\mathbb{Z}$-{\it word} over the alphabet $\Si$ is a {\it biinfinite word} : 
$$ \ldots x(-3).x(-2).x(-1).x(0).x(1).x(2).x(3) \ldots $$
\noi where for each $i \in \mathbb{Z}$, the letter $x(i)$ is in the alphabet $\Si$. 
\nl The {\it set of } $\mathbb{Z}$-{\it words} over  the alphabet $\Si$ is denoted by $\Si^\mathbb{Z}$. It can be identified with the set of functions 
from $\mathbb{Z}$ into $\Si$. 

\hs  We now define  one-dimensional cellular automata. 

\begin{Deff}
\noi 
\begin{enumerate}
\ite  A one-dimensional cellular automaton $\mathcal{A}$ is a pair $(Q, \delta)$, where $Q$ is a finite set of states and $\delta$ is a function  from 
$Q^3$ into $Q$, called the ``local transition function" of the automaton.
\ite A configuration of the cellular automaton $\mathcal{A}$ is a mapping from $\mathbb{Z}$ into $Q$, i.e. an element of $Q^\mathbb{Z}$. 
\ite The global transition function of the cellular automaton $\mathcal{A}$ is the function $\Delta$ from $Q^\mathbb{Z}$ into $Q^\mathbb{Z}$ defined 
by : 
$$\fa z \in \mathbb{Z}, ~ \Delta(C)(z) = \delta ( C(z-1), C(z), C(z+1) )$$ 
\noi So the cellular automaton may be also denoted by $(Q^\mathbb{Z}, \Delta)$. 
\end{enumerate}
\end{Deff}

\noi Notice that we have defined   the  ``local transition function" of a   one-dimensional cellular      automaton as a function  from 
$Q^3$ into $Q$, i.e. the next state of a given cell depends only on the given  state of this cell and of its closest neighbours. However all the results in this paper 
can be easily generalized to the case of a  ``local transition function"  being  a function from $Q^{2r+1}$ into $Q$, for an integer $r\geq 1$, 
as in \cite{Sutner-JCA}.

\hs  
We  recall now the notion of B\"uchi automaton reading 
infinite words over a finite alphabet, which can be found for instance in \cite{Thomas90,Staiger97}.

\begin{Deff}
A   B\"uchi automaton 
 is a sextuple $\mathcal{A}=(K, \Si, \Delta, q_0, F)$, where 
$K$ is a finite set of states, $\Si$ is a  finite  alphabet, 
$\Delta \subseteq K \times \Si   \times K$ is 
the set of transitions, $q_0 \in K$ is the initial state,  and $F \subseteq K$ is the set of 
accepting states. 
\nl A run  $r$ of the  
 B\"uchi automaton $\mathcal{A}$ on an $\om$-word $x \in \Sio$  is an infinite sequence of states $(q_i)_{i \geq 0}$ such 
that for each integer $i\geq 0$ ~~$(q_i, x(i), q_{i+1}) \in \Delta$. Notice that the first state of the run is the initial state $q_0$. 
\nl The run  is said to be successful iff there exists a final state $q_f \in F$ 
and infinitely many integers $i\geq 0$ such that $q_i=q_f$. 
\nl The $\om$-language  $L(\mathcal{A})\subseteq \Sio$  accepted by the B\"uchi automaton $\mathcal{A}$ is 
the set of  $\om$-words $x$ such that there is a successful run $r$ of  $\mathcal{A}$ on $x$. 
\nl An $\om$-language $L \subseteq \Sio$ is a regular $\om$-language iff there is a B\"uchi automaton $\mathcal{A}$ such that 
$L = L(\mathcal{A})$. 
\end{Deff}

\noi Notice that one can  consider a relation 
$R \subseteq \Si_1^\om \times \Si_2^\om  \times \ldots \times \Si_n^\om$, where $\Si_1, \Si_2, \ldots \Si_n$, are  finite alphabets, 
as an $\om$-language over the product alphabet $\Si_1 \times \Si_2  \times \ldots \times \Si_n$. 

\hs We now recall some fundamental closure properties of regular $\om$-languages. 

\begin{The}[see \cite{Thomas90,PerrinPin}]
The class of regular  $\om$-languages is effectively closed under finite union, finite intersection, and complementation, i.e. 
we can effectively construct, from two B\"uchi automata    $\mathcal{A}$ and $\mathcal{B}$, some 
B\"uchi automata 
$\mathcal{C}_1$, $\mathcal{C}_2$, and $\mathcal{C}_3$,  such that $L(\mathcal{C}_1)=L(\mathcal{A}) \cup L(\mathcal{B})$, 
$L(\mathcal{C}_2)=L(\mathcal{A}) \cap L(\mathcal{B})$,  and $L(\mathcal{C}_3)$ is the complement of $L(\mathcal{A})$. 
\end{The}

\section{$\om$-automatic structures}

\noi  Let now $\mathcal{M}=(M, (R_i^M)_{1\leq i\leq n})$  be  a relational structure, 
  where $M$ is the domain  and,  for each $i\in [1, n]$, ~ $R_i^M$ is a relation 
of finite arity $n_i$ on the domain $M$. The structure is said to be  $\om$-automatic 
if  the domain and the relations on the domain are accepted by B\"uchi automata  in the following sense.

\begin{Deff}[see \cite{Hodgson,Blumensath99}]
Let $\mathcal{M}=(M, (R_i^M)_{1\leq i\leq n})$ be a relational structure, where $n\geq 1$ is an integer,  and each relation $R_i$ is of finite arity $n_i$. 
\nl An  $\om$-automatic presentation of the structure $\mathcal{M}$  is formed by   a tuple of   B\"uchi   automata 
$(\mathcal{A}, (\mathcal{A}_i)_{1\leq i\leq n})$,  and a mapping $h$ from $L(\mathcal{A})$ onto $M$,  such that :   
\begin{enumerate}
\ite $L(\mathcal{A})\subseteq \Si^\om$,  for a finite alphabet $\Si$, and 
\ite 
For each $i \in [1, n]$, the automaton $\mathcal{A}_i$ accepts an $n_i$-ary relation $R'_i$ on 
$L(\mathcal{A})$,  and 
\ite   The mapping $h$ is an isomorphism from the  structure $( L(\mathcal{A}),  (R'_i)_{1 \leq i \leq n})$ onto the  structure 
$\mathcal{M}=(M, (R_i^M)_{1\leq i\leq n})$. 
\end{enumerate}
\noi A relational structure is said to be  $\om$-automatic if it has an  $\om$-automatic presentation. 
\end{Deff} 

\begin{Rem}  
An  $\om$-automatic presentation of a structure as defined above is often said  to be   an  {\it injective}  $\om$-automatic presentation. 
It is actually a particular case of the more general notion of (non injective) $\om$-automatic presentation of a structure, see \cite{HjorthKMN08}. 
However this restricted notion will be sufficient in the sequel so we have only given the definition of an $\om$-automatic presentation of a structure in this particular 
case. The reader can see \cite{KuskeLohrey,HjorthKMN08} for more information on this subject. 
\end{Rem}

\noi  We recall now two important properties of $\om$-automatic structures. 

\begin{The}[Hodgson  \cite{Hodgson}]\label{dec}
The first-order theory of an $\om$-automatic structure is decidable. 
\end{The}

\begin{The}[see \cite{Blumensath99}]
The class of  $\om$-automatic structures is closed under first-order interpretations. If 
$\mathcal{M}$ is an $\om$-automatic structure and $\mathcal{M}'$ is a relational structure which is 
first-order interpretable in the structure $\mathcal{M}$, then the structure  $\mathcal{M}'$ is also $\om$-automatic.  
\end{The}

\noi  Some examples of  $\om$-automatic structures can be found in 
\cite{BlumensathGraedel04,KuskeLohrey}. 

\section{Decidability properties}

\noi We consider a cellular automaton $\mathcal{A}=(Q^\mathbb{Z}, \Delta)$, where $Q^\mathbb{Z}$ is the set of all possible configurations and 
$\Delta$ is the global transition function. 
\nl We can replace the function $\Delta$ by its graph, which is a binary relation $\longrightarrow$ on the space $Q^\mathbb{Z}$. 
The space of configurations of  the  cellular automaton $\mathcal{A}$, equipped with the ``next configuration relation" $\longrightarrow$  will be denoted 
$$\mathcal{S}_\mathcal{A}=(Q^\mathbb{Z}, \longrightarrow).$$ 

\noi It is a simple relational structure with only one binary relation on the domain $Q^\mathbb{Z}$.  Sutner proved that, for each cellular automaton 
$\mathcal{A}$,  the first-order theory of the 
structure $\mathcal{S}_\mathcal{A}$ is decidable, \cite{Sutner-JCA}. We shall see that this result can be deduced from Theorem \ref{dec} on $\om$-automatic structures. 

\begin{Lem}\label{aut}
Let $\mathcal{A}=(Q^\mathbb{Z}, \Delta)$ be a one-dimensional cellular automaton. 
Then  the structure $\mathcal{S}_\mathcal{A}=(Q^\mathbb{Z}, \longrightarrow)$
is  $\om$-automatic. 
\end{Lem}

\proo We can represent a configuration $c\in Q^\mathbb{Z}$ by an $\om$-word $x_c$ over the alphabet $Q \times Q$ which is simply defined by 
$x_c(0)=(c(0), c(0))$, and,  for each integer $n\geq 1$,  $x_c(n)=(c(-n), c(n))$. 
\nl The set of $\om$-words in the form $x_c$ for some configuration $c\in Q^\mathbb{Z}$ is simply the set of $\om$-words $x$ in $(Q \times Q)^\om$  
such that $x(0)=(a, a)$ for some $a\in Q$. It is clearly a regular $\om$-language accepted by a B\"uchi automaton $\mathcal{B}$ and the function 
$h: x_c \ra c$ is a bijection from $L(\mathcal{B})$ onto $Q^\mathbb{Z}$. 

\hs Due to the {\bf local } definition of the {\bf ``local transition function"} of  a cellular  automaton, it is easy to see that the  set of  pairs of 
$\om$-words  $(x_c, x_{c'})$ such that $c \longrightarrow c'$ is accepted by a B\"uchi automaton $\mathcal{B}_1$ reading words over the product alphabet 
$(Q \times Q) \times (Q \times Q)$. 
\nl In fact  the relation $c \longrightarrow c'$ can be checked by a scanner, as described in \cite[page 283]{PerrinPin},  i.e. 
``a  machine equipped with a finite memory 
and a sliding window", here of fixed size $3$,  to scan the input $\om$-word  $(x_c, x_{c'})$. 
\ep 

\hs Notice that Sutner proved in a similar way that, when representing configurations by {\it biinfinite words}, the relation $c \longrightarrow c'$ is accepted by a 
B\"uchi automaton reading {\it biinfinite words}, see \cite{PerrinPin,Sutner-JCA} for more details. Then he proved that the first-order theory of the structure 
$\mathcal{S}_\mathcal{A}$ is decidable. We can obtain now this result as a consequence of Lemma \ref{aut} and Theorem \ref{dec}. 

\begin{Cor}\label{cor}
Let $\mathcal{A}=(Q^\mathbb{Z}, \Delta)$ be a one-dimensional cellular automaton. 
Then  the first-order theory of the structure 
$\mathcal{S}_\mathcal{A}$ is decidable. 
\end{Cor}

\proo It follows directly from  Lemma \ref{aut} and the fact that the first-order theory of an $\om$-automatic structure is decidable, proved by Hodgson in 
\cite{Hodgson}.
\ep

\hs Notice that if $R_1, R_2, \ldots , R_n$ are  first-order definable relations over $Q^\mathbb{Z}$ then the expanded structure 
$(Q^\mathbb{Z}, \longrightarrow, R_1, R_2, \ldots , R_n)$ is also $\om$-automatic and then has a first-order decidable theory. 

\hs We have seen that Corollary \ref{cor} is an easy consequence of the fact that the structure $\mathcal{S}_\mathcal{A}$ is $\om$-automatic and 
of Hodgson's result on the decidability of the first-order theory of an  $\om$-automatic structure \cite{Hodgson}. 
\nl On the other hand, an extension of Hodgson's result to first-order logic extended with the infinity quantifier $\exists^\infty$ 
has been proved by Blumensath and Gra\"edel in \cite{BlumensathGraedel04}, where   
$\exists^\infty$ means ``there are infinitely many ". 
Other extensions of Theorem \ref{dec} involving other counting and cardinality quantifiers have 
been proved by  Kuske and Lohrey proved in \cite{KuskeLohrey}. 
We   now recall the definition of these  quantifiers and the results for (injectively)  $\om$-automatic structures, \cite{KuskeLohrey}. 

\hs  The quantifier $\exists$ is usual in first-order logic. In addition we can consider the infinity quantifier $\exists^\infty$, the cardinality quantifiers 
$\exists^\kappa$ for any cardinal $\kappa$, and the modulo quantifiers $\exists^{(t, k)}$ where $t, k$ are integers such that  $0\leq t<k$. 
The set of first-order formulas is denoted by $\mathrm{FO}$. For a class of cardinals $C$, the set of formulas using the additional quantifiers 
$\exists^\infty$, $\exists^\kappa$ for  cardinals $\kappa \in C$, and modulo quantifiers $\exists^{(t, k)}$, is denoted by 
 $\mathrm{FO}(\exists^\infty, (\exists^\kappa)_{\kappa \in C}, (\exists^{(t, k)})_{0\leq t<k})$. 
\nl If $\mathcal{L}$ is a set of formulas, the $\mathcal{L}$-theory of a structure $\mathcal{M}$ is the set of sentences (i.e., formulas without free variables) 
in $\mathcal{L}$ 
that hold in $\mathcal{M}$. 

\hs 
If  $\mathcal{M}$ is a relational structure in a signature $\tau$ which contains only relational 
symbols and equality, and $\psi$ is a formula in   $\mathrm{FO}(\exists^\infty, (\exists^\kappa)_{\kappa \in C}, (\exists^{(t, k)})_{0\leq t<k})$, then 
the semantics of the above  quantifiers are defined as follows. 

\begin{itemize}
\ite $\mathcal{M} \models \exists^\infty x ~ \psi$  if and only if there are infinitely many $a\in \mathcal{M}$ such that $\mathcal{M} \models  \psi(a)$.
\ite $\mathcal{M} \models \exists^\kappa x ~ \psi$ if and only if the set $\psi^\mathcal{M}=\{a \in \mathcal{M} \mid \mathcal{M} \models  \psi(a) \}$ 
has cardinality $\kappa$.
\ite  $\mathcal{M} \models \exists^{(t, k)} x  ~\psi$ if and only if the set $\psi^\mathcal{M}=\{a \in \mathcal{M} \mid \mathcal{M} \models  \psi(a) \}$
 is finite and its cardinal $|\psi^\mathcal{M}|$ is equal to $t$ modulo $k$. 
\end{itemize}

\noi The $\mathrm{FO}(\exists^\infty)$-theory of an (injectively) $\om$-automatic structure has been shown to be decidable by 
Blumensath and Gra\"edel in \cite{BlumensathGraedel04}.  
We are now going to recall the more general result proved by Kuske and Lohrey  in \cite{KuskeLohrey}. 

\begin{The}[\cite{KuskeLohrey}]\label{fund}
Let an (injective) $\om$-automatic presentation of a structure $\mathcal{M}$   formed by   a tuple of   B\"uchi   automata 
$(\mathcal{A}, (\mathcal{A}_i)_{1\leq i\leq n})$,  and a mapping $h$. Let $C$ be an at most countably infinite set of cardinals and 
$\psi(x_1, \ldots, x_n)$ be a formula of $\mathrm{FO}(\exists^\infty, (\exists^\kappa)_{\kappa \in C}, (\exists^{(t, k)})_{0\leq t<k})$ over the signature 
of the structure $\mathcal{M}$. Then the relation 
$$R = \{ (u_1, \ldots , u_n) \in L(\mathcal{A})^n \mid \mathcal{M} \models \psi(h(u_1), \ldots , h(u_n)) \}$$
\noi is effectively $\om$-automatic, i.e., one can effectively construct a B\"uchi automaton $\mathcal{A}_R$ such that 
$L(\mathcal{A}_R)=R$. 
\end{The}

\begin{Cor}[\cite{KuskeLohrey}]
Let $\mathcal{M}$ be an (injectively) $\om$-automatic structure and let $C$ be an at most countably infinite set of cardinals. Then the structure $\mathcal{M}$ 
has a decidable $\mathrm{FO}(\exists^\infty, (\exists^\kappa)_{\kappa \in C}, (\exists^{(t, k)})_{0\leq t<k})$-theory.
\end{Cor}

\noi We can now 
apply these new results to the study of cellular automata. 

\begin{Cor}
Let $\mathcal{A}=(Q^\mathbb{Z}, \Delta)$ be a one-dimensional cellular automaton 
and let $C$ be an at most countably infinite set of cardinals. Then the structure 
$\mathcal{S}_\mathcal{A}=(Q^\mathbb{Z}, \longrightarrow)$
 has a decidable $\mathrm{FO}(\exists^\infty, (\exists^\kappa)_{\kappa \in C}, (\exists^{(t, k)})_{0\leq t<k})$-theory. 
\end{Cor}

\noi Then Sutner's result is extended to a more powerful logic, containing counting and cardinality quantifiers. 

\hs We now give some examples of  decidable properties we get  for one-dimensional 
cellular automata. 

\hs Applying Theorem \ref{fund} and its corollaries, 
we can for example decide whether there are finitely many (respectively, countably many, uncountably many)
configurations having  finitely many (respectively, countably many, uncountably many) preimages by the global transition function $\Delta$ of a cellular 
automaton $\mathcal{A}$. 
Moreover we can construct a B\"uchi automaton accepting exactly the (set of $\om$-words representing the) configurations having 
finitely many (respectively, countably many, uncountably many) preimages. 

\hs We can also determine exactly the cardinal of the set of fixed points.  Moreover if $R$ is a regular $\om$-language containing 
$\om$-words representing configurations, then we can determine the cardinal of fixed points  $c \in Q^\mathbb{Z}$ such that $x_c \in R$. 
We can also construct a B\"uchi automaton accepting exactly the (set of $\om$-words $x_c$ representing the) fixed points $c$ such that $x_c\in R$. 

\hs In a similar way we can determine whether there are infinitely many (respectively, countably many, uncountably many) $3$-cycles. If there are only finitely 
many $3$-cycles then we can compute the cardinal of this finite set. And we can construct a B\"uchi 
automaton accepting exactly $\om$-words $(x_c, x_{c'}, x_{c''})$ for the  $3$-cycles $(c, c', c'')$.

\section{More efficient algorithms}

\noi A problem in the concrete application of Theorem \ref{fund} and its corollaries  is the complexity of the algorithms we get from them. 
Sutner noticed  in \cite{Sutner-JCA} that the algorithm he got for the decidability of the first-order theory  
of the phase-space  $\mathcal{S}_\mathcal{A}$ of a cellular automaton $\mathcal{A}$  is 
in  general non elementary. This is also the case for the algorithms obtained for (injectively) $\om$-automatic structures, see \cite{KuskeLohrey}.

\hs However some more efficient algorithms can be obtained from results of Kuske and Lohrey on $\om$-automatic structures of bounded degree. 
A relational structure is of bounded degree iff its Gaifman-graph has bounded degree. The 
Gaifman-graph of the structure $\mathcal{S}_\mathcal{A}=(Q^\mathbb{Z}, \longrightarrow)$ is simply the undirected graph 
$(Q^\mathbb{Z}, R)$, where for any configurations $c, c' \in Q^\mathbb{Z}$, 
[ $R(c, c')$ ] iff [ $c \longrightarrow c'$ or $c' \longrightarrow c$ ], (see the precise definition of the Gaifman-graph  in \cite{KuskeLohrey}). 
\nl Notice first that, for a   cellular automaton $\mathcal{A}=(Q^\mathbb{Z}, \Delta)$, any configuration $c\in  Q^\mathbb{Z}$ has a unique successor 
for the relation $\longrightarrow$. Then the phase-space $\mathcal{S}_\mathcal{A}$ is a structure of bounded degree if and only if there is an integer $k\geq 1$ 
such that any configuration $c \in Q^\mathbb{Z}$ has at most $k$ pre-images  by  the global transition function $\Delta$ of the cellular automaton 
$\mathcal{A}$. 

\hs On the other hand it is well known that this property corresponds to the important class of {\it surjective} cellular automata, 
see for example \cite[page 217]{Kurka}. As this class is an important one, we are going to recall results of Kuske and Lohrey on 
 $\om$-automatic structures of bounded degree, see \cite{KuskeLohrey} for more details. 

\hs Kuske and Lohrey considered  the logic $\mathcal{L}(\mathcal{Q}_u)$ in a relational signature $\tau$. 
Formulas of the logic $\mathcal{L}(\mathcal{Q}_u)$ are built from atomic formulas of the form $R(x_1, \ldots, x_k)$ where $R \in \tau$ is a 
$k$-ary relational symbol and $x_1, \ldots, x_k$ are first-order variables ranging over the universe of the underlying structure, using boolean connectives and 
quantifications of the form 
$$\mathcal{Q}_\mathcal{C}y: (\psi_1(\bar{x}, y), \ldots,     \psi_n(\bar{x}, y) ).$$
\noi Above,  each $\psi_i$ is already a formula of $\mathcal{L}(\mathcal{Q}_u)$, $\bar{x}$ is a sequence of variables, 
and $\mathcal{C}$ is an $n$-ary relation 
over cardinals. 
\nl We now recall the semantics of a $\mathcal{Q}_\mathcal{C}$ quantifier. 
Let   $\mathcal{M}=(M, R_1, \ldots, R_j)$ be a $\tau$-structure where $M$ is the domain 
and each $R_i$ is a $n_i$-ary relation on $M$. Let  $\bar{u}$ be a tuple of elements of $M$ of the same length as $\bar{x}$, then 
$\mathcal{M} \models \mathcal{Q}_\mathcal{C}y: (\psi_1(\bar{u}, y), \ldots,     \psi_n(\bar{u}, y) )$ if and only if $(c_1, \ldots, c_n) \in \mathcal{C}$, where 
each $c_i$ is the cardinality of the set $\{ a \in M \mid  \mathcal{M} \models \psi_i(\bar{u}, a) \}$.

\hs The quantifier  $\mathcal{Q}_\mathcal{C}$ is called an $n$-dimensional counting quantifier. 
\nl Notice that all the quantifiers $\exists$, $\exists^\infty$, $\exists^\kappa$, for a cardinal $\kappa$, and the modulo quantifiers $\exists^{(t, k)}$, 
are particular cases of one-dimensional counting quantifiers. A well known two-dimensional counting quantifier is the H\"artig quantifier 
$I y  :  (\psi_1(\bar{x}, y), \psi_2(\bar{x}, y) )$ expressing ``the number of $y$ satisfying $\psi_1(\bar{x}, y)$ is equal to 
the number of $y$ satisfying $\psi_2(\bar{x}, y)$"  (the relation $\mathcal{C}$  is here the identity relation on cardinals). 

\hs Let $\mathbb{C}$ be a class of relations on cardinals. A formula  in $\mathcal{L}(\mathbb{C})$ is a formula of    $\mathcal{L}(\mathcal{Q}_u)$
 which uses only quantifiers of the form 
$\mathcal{Q}_\mathcal{C}$ for $\mathcal{C} \in \mathbb{C}$.  The logic $\mathrm{FO}(\mathbb{C})$ is the first order logic extended with the 
quantifiers $\mathcal{Q}_\mathcal{C}$ for $\mathcal{C} \in \mathbb{C}$. If $\mathbb{C}=\{\mathcal{C}\}$ is a singleton then 
we write simply $\mathrm{FO}(\mathcal{C})$ 
instead of $\mathrm{FO}(\mathbb{C})$. 

\hs We now recall the decidability result of Kuske and Lohrey. 

\begin{The}[Kuske and Lohrey   \cite{KuskeLohrey}]
Let $\mathbb{C}=\{ \mathcal{C}_i \mid i\in \mathbb{N} \}$, where $\mathcal{C}_i$ is a relation on $\mathbb{N} \cup \{ \aleph_0, 2^{\aleph_0}\}$. 
Let $\mathcal{M}$ be an (injectively) $\om$-automatic structure of bounded degree. Then the $\mathcal{L}(\mathbb{C})$-theory of 
$\mathcal{M}$  can be decided in triply exponentional space by a 
Turing machine with oracle $\{ (i, \bar{c}) \mid i \in \mathbb{N}, \bar{c}\in \mathcal{C}_i \}$.
\end{The}

\begin{Cor}
Let $\mathcal{A}$ be a surjective one-dimensional cellular automaton,  
and let $\mathbb{C}=\{ \mathcal{C}_i \mid i\in \mathbb{N} \}$, where $\mathcal{C}_i$ is a relation on $\mathbb{N} \cup \{ \aleph_0, 2^{\aleph_0}\}$. 
 Then the $\mathcal{L}(\mathbb{C})$-theory of the structure  $\mathcal{S}_\mathcal{A}$ 
 can be decided in triply exponentional space by a 
Turing machine with oracle $\{ (i, \bar{c}) \mid i \in \mathbb{N}, \bar{c}\in \mathcal{C}_i \}$.
\end{Cor}

\noi Notice that it is proved in \cite{KuskeLohrey} that even if the continuum hypothesis is not satisfied then allowing cardinals $\kappa$ with 
$\aleph_0 < \kappa < 2^{\aleph_0}$ in the relations $\mathcal{C}_i$ ``does not change the results on $\om$-automatic structures". 

\hs Notice that in fact the logic $\mathcal{L}(\mathcal{Q}_u)$  has a restricted expressive power for structures of bounded degree. This follows from 
the following result. 

\begin{The}
Let $\mathcal{M}$ be a $\tau$-structure of bounded degree, and let $\phi(\bar{x})$ be a formula of  $\mathcal{L}(\mathcal{Q}_u)$. Then there is a 
first-order formula $\psi(\bar{x}) \in \mathrm{FO}$ such that $\mathcal{M} \models \fa \bar{x} ( \phi(\bar{x}) \leftrightarrow \psi(\bar{x}) )$. 
\end{The}

\noi In the case of (injectively) $\om$-automatic structure of bounded degree we can effectively construct the formula $\psi$, see 
\cite[Corollary 3.13]{KuskeLohrey}. 
We give now directly the important corollary obtained in the case of surjective one-dimensional cellular automata. 

\begin{Cor}
Let $\mathcal{A}$ be a surjective one-dimensional cellular automaton, 
and let $\mathbb{C}=\{ \mathcal{C}_i \mid i\in \mathbb{N} \}$, where $\mathcal{C}_i$ is a relation on $\mathbb{N} \cup \{ \aleph_0, 2^{\aleph_0}\}$. 
For a given formula $\phi(\bar{x})$ $\in \mathcal{L}(\mathbb{C})$, one can construct in elementary space (modulo $\mathbb{C}$) a first-order formula 
$\psi(\bar{x})$ and a  B\"uchi automaton $\mathcal{B}_\phi$ such that 
for any $\bar{u}=(c_1, \ldots , c_k)  \in (Q^\mathbb{Z})^k$,  (where $k=|\bar{x}|$) ,
$$\mathcal{S}_\mathcal{A} \models \phi(\bar{u})  \Leftrightarrow 
 \mathcal{S}_\mathcal{A} \models \psi(\bar{u})   \Leftrightarrow (x_{c_1}, \ldots,  x_{c_k}) \in L(\mathcal{B_\phi}).$$
\end{Cor}

\section{From cellular automata to automatic structures}

\noi We have seen  that we can infer new decidability results for cellular automata from  recent results about $\om$-automatic structures. 
\nl On the other hand we can also obtain new results about automatic structures from results about cellular automata. 

\hs If $\mathcal{A}=(Q, \delta)$ is a one-dimensional cellular automaton  with a particular quiescient state $\# \in Q$, 
 we consider  the set of configurations with finite support $\mathcal{C}_{fin}$ of $\mathcal{A}$. These configurations are in the form 
$c\in Q^\mathbb{Z}$ with  $c(n)=\#$ for all $n \leq -k $ or $n \geq k$ where $k\geq 0$ is an integer. 

\hs We shall consider in this section the notion of automatic structure which is defined as in the case of an $\om$-automatic structure but using 
automata reading {\it finite words}  instead of {\it infinite words}. Thus an automatic structure is always a countable structure, 
see \cite{BlumensathGraedel04,KNRS,Hodgson} for more details. 
We can now state the following result. 

\begin{Pro}
Let $\mathcal{A}=(Q^\mathbb{Z}, \Delta)$ be a one-dimensional cellular automaton. 
Then  the structure $\mathcal{S}_{\mathcal{A}, fin}=(\mathcal{C}_{fin}, \longrightarrow)$
is  automatic. 
\end{Pro}

\noi Then we have new examples of directed automatic graphs.  
The directed graph $\mathcal{S}_{\mathcal{A}, fin}$ of the set of configurations with finite support $\mathcal{C}_{fin}$ of a given 
one-dimensional cellular automaton $\mathcal{A}$, equipped with the ``next configuration relation $\longrightarrow$". 

\hs In a directed graph  $(G, \longrightarrow)$, the reachability relation $\stackrel{\star}{\longrightarrow}$ is the reflexive and transitive closure of the 
relation $\longrightarrow$. It is clear that in a recursive graph, hence also in an automatic graph, the reachability relation is recursively enumerable. 
Sutner proved that actually for each recursively enumerable Turing degree ${\bf d}$ there is a cellular automaton $\mathcal{A}$ such that the 
reachability relation in the graph $\mathcal{S}_{\mathcal{A}, fin}$ has exactly the Turing degree ${\bf d}$, see \cite{Sutner2}. 
So we obtain the following corollary. 

\begin{Cor}\label{deg-ca}
For each recursively enumerable Turing degree ${\bf d}$ there is an automatic graph $(G, \longrightarrow)$ in which the 
reachability relation  has exactly the Turing degree ${\bf d}$. 
\end{Cor}

\noi Recall that the structure of Turing degrees is very complicated. If ${\bf 0}$ denoted the degree of recursive sets and ${\bf 0'}$ denotes the 
degree of complete recursively enumerable sets, then by the famous result of Friedberg and Muchnik answering Post's question, there is a Turing degree 
${\bf d}$ such that ${\bf 0}< {\bf d}< {\bf 0'}$, see \cite{rog}.  Moreover by Sack's Density Theorem, for two  recursively enumerable degrees 
${\bf d_1}<{\bf d_2}$ there is a third one ${\bf d_3}$ such that ${\bf d_1}<{\bf d_3}<{\bf d_2}$. 

\hs The above Corollary \ref{deg-ca} can be refined. In a directed graph $(G, \longrightarrow)$ the {\it confluence relation} $Conf_G$ is defined by: 
for any $x, y \in G$, $Conf_G(x,y)$ iff there is some $z\in G$ such that $x \stackrel{\star}{\longrightarrow} z$ and $y  \stackrel{\star}{\longrightarrow} z$. 
Sutner proved that for any recursively enumerable degrees ${\bf d_1}$ and ${\bf d_2}$ there is a a cellular automaton $\mathcal{A}$ such that the 
reachability relation in the graph $\mathcal{S}_{\mathcal{A}, fin}$ has exactly the Turing degree ${\bf d_1}$ and the confluence relation in the graph 
 $\mathcal{S}_{\mathcal{A}, fin}$ has exactly the Turing degree ${\bf d_2}$. So we can state the following corollary. 

\begin{Cor}\label{deg-ca2}
For each recursively enumerable Turing degrees ${\bf d_1}$ and  ${\bf d_2}$  there is an automatic graph $(G, \longrightarrow)$ in which the 
reachability relation  has exactly the Turing degree ${\bf d_1}$ and  the confluence relation has exactly the Turing degree ${\bf d_2}$. 
\end{Cor}

\noi 
Khoussainov, Nies, Rubin, and Stephan proved in \cite{KNRS} that the isomorphism problem for automatic graphs is $\Si_1^1$-complete, 
hence as complicated as the  isomorphism problem for recursive  graphs. This showed that the class of  automatic graphs is very rich. The above 
results confirm this from the point of view of Turing degrees. 

\hs Notice that if an automatic graph is viewed as the set of configurations of  an infinite state transition system, then decision problems naturally appear and are 
very important in this context. 
Some  decision problems for automatic graphs have been recently studied by Kuske and Lohrey in \cite{KuskeLohrey2}. For instance they proved that 
the problem of the existence of   an Hamiltonian path in a planar automatic graph of bounded degree is $\Si_1^1$-complete, and that the 
problem of the existence of   an Euler path in an automatic graph is $\Pi_2^0$-complete. 

\hs We can now infer other results from some results of Sutner on cellular automata. 
Let $\mathbb{C}_{{\bf d}}$ be the class of cellular automata $\mathcal{A}$
such that the  reachability relation of the graph $\mathcal{S}_{\mathcal{A}, fin}$  has exactly degree ${\bf d}$. 
Sutner proved that it is $\Si_3^{{\bf d}}$-complete to decide membership of a given cellular automaton in the class $\mathbb{C}_{{\bf d}}$, for 
any recursively enumerable Turing degree ${\bf d}$. In particular,  it is $\Si_3^0$-complete to decide membership in  $\mathbb{C}_{{\bf 0}}$ and it is 
$\Si_4^0$-complete to decide membership in  $\mathbb{C}_{{\bf 0'}}$, see \cite{Sutner2}. We can now state  the following result on automatic graphs. 

\begin{Cor}\label{deg-ca3}
\noi 
\begin{enumerate}
\ite It is $\Si_3^0$-complete to decide  whether the reachability relation of a given automatic graph is recursive. 
\ite It is $\Si_4^0$-complete to decide  whether the reachability relation of a given automatic graph is $\Si_1^0$-complete. 
\end{enumerate}
\end{Cor}

\proo  The lower bound results follow  from the correponding results for graphs of  cellular automata. 
\nl The fact that deciding whether the reachability relation of a given automatic graph is recursive is in the class $\Si_3^0$ follow from the known fact 
that   deciding  whether a given recursively enumerable set is recursive is in the class $\Si_3^0$, see \cite[page 327]{rog}. 
 The fact that deciding whether the reachability relation of a given automatic graph is $\Si_1^0$-complete is in the class $\Si_4^0$ follow from the known fact 
that  deciding whether a given recursively enumerable set is $\Si_1^0$-complete is in the class $\Si_4^0$, see \cite[page 330]{rog}. 
\ep 

\section{Concluding remarks}

\noi We have obtained  some new decidability results about cellular automata, showing that their graphs are $\om$-automatic
and applying recent results of Kuske and Lohrey. 
\nl  On the other hand we have got new examples of automatic graphs and new results about these graphs, applying recent results of Sutner 
about cellular automata. 
\nl We hope that the interplay of the two domains of cellular automata and of automatic structures will be fruitful and will provide  new results 
in both fields of research.


\begin{thebibliography}{HKMN08}

\bibitem[AP72]{AmorosoPatt}
S.~Amoroso and Y.~N. Patt.
\newblock Decision procedures for surjectivity and injectivity of parallel maps
  for tesselation structures.
\newblock {\em Journal of Computer and Systems Science}, 6:448--464, 1972.

\bibitem[BG04]{BlumensathGraedel04}
A.~Blumensath and E.~Gr{\"a}del.
\newblock Finite presentations of infinite structures: Automata and
  interpretations.
\newblock {\em Theory of Computing Systems}, 37(6):641--674, 2004.

\bibitem[Blu99]{Blumensath99}
A.~Blumensath.
\newblock {\em Automatic Structures}.
\newblock Diploma Thesis, RWTH Aachen, 1999.

\bibitem[DM94]{Delorme-Mazoyer}
M.~Delorme and J.~Mazoyer, editors.
\newblock {\em Cellular Automata, a parallel Model}.
\newblock Kluwer Academic, 1994.

\bibitem[HKMN08]{HjorthKMN08}
G.~Hjorth, B.~Khoussainov, A.~Montalb{\'a}n, and A.~Nies.
\newblock From automatic structures to {B}orel structures.
\newblock In {\em Proceedings of the Twenty-Third Annual IEEE Symposium on
  Logic in Computer Science, LICS 2008, 24-27 June 2008, Pittsburgh, PA, USA},
  pages 431--441. IEEE Computer Society, 2008.

\bibitem[Hod83]{Hodgson}
B.~R. Hodgson.
\newblock D\'ecidabilit\'e par automate fini.
\newblock {\em Annales Scientifiques de Mathématiques du Québec}, 7(1):39--57,
  1983.

\bibitem[KL08]{KuskeLohrey}
D.~Kuske and M.~Lohrey.
\newblock First-order and counting theories of omega-automatic structures.
\newblock {\em Journal of Symbolic Logic}, 73(1):129--150, 2008.

\bibitem[KL09]{KuskeLohrey2}
D.~Kuske and M.~Lohrey.
\newblock Some natural problems in automatic graphs.
\newblock {\em Journal of Symbolic Logic}, 2009, to appear.
\newblock Available from http://www.informatik.uni-leipzig.de/~kuske/.

\bibitem[KNRS07]{KNRS}
B.~Khoussainov, A.~Nies, S.~Rubin, and F.~Stephan.
\newblock Automatic structures: Richness and limitations.
\newblock {\em Logical Methods in Computer Science}, 3(2):1--18, 2007.

\bibitem[Kur03]{Kurka}
P.~Kurka.
\newblock {\em Topological and Symbolic Dynamics}, volume~11 of {\em Cours
  Sp\'ecialis\'es}.
\newblock Soci\'et\'e Math\'ematique de France, 2003.

\bibitem[PP04]{PerrinPin}
D.~Perrin and J.-E. Pin.
\newblock {\em Infinite words, automata, semigroups, logic and games}, volume
  141 of {\em Pure and Applied Mathematics}.
\newblock Elsevier, 2004.

\bibitem[Rog67]{rog}
H.~Rogers.
\newblock {\em Theory of Recursive Functions and Effective Computability}.
\newblock McGraw-Hill, New York, 1967.

\bibitem[Sta97]{Staiger97}
L.~Staiger.
\newblock $\omega$-languages.
\newblock In {\em Handbook of formal languages, Vol.\ 3}, pages 339--387.
  Springer, Berlin, 1997.

\bibitem[Sut08a]{Sutner2}
K.~Sutner.
\newblock Classification of cellular automata.
\newblock In {\em Encyclopedia of Complexity and System Science}. 2008.
\newblock to appear, Available from http://www.cs.cmu.edu/~sutner/.

\bibitem[Sut09]{Sutner-JCA}
K.~Sutner.
\newblock Model checking one-dimensional cellular automata.
\newblock {\em Journal of Cellular Automata}, 4(3):213--224, 2009.

\bibitem[Tho90]{Thomas90}
W.~Thomas.
\newblock Automata on infinite objects.
\newblock In J.~van Leeuwen, editor, {\em Handbook of Theoretical Computer
  Science}, volume B, Formal models and semantics, pages 135--191. Elsevier,
  1990.

\bibitem[Wol86]{Wolfram}
S.~Wolfram, editor.
\newblock {\em Theory and Applications of Cellular Automata}.
\newblock World Scientific Press, 1986.

\end{thebibliography}
\end{document}